\documentclass[twocolumn]{article}

\usepackage{arxiv}

\usepackage[all]{nowidow}

\usepackage[utf8]{inputenc} % allow utf-8 input
\usepackage[T1]{fontenc}    % use 8-bit T1 fonts
\usepackage{hyperref}       % hyperlinks
\usepackage{url}            % simple URL typesetting
\usepackage{booktabs}       % professional-quality tables
\usepackage{amsfonts}       % blackboard math symbols
\usepackage{nicefrac}       % compact symbols for 1/2, etc.
\usepackage{microtype}      % microtypography

\usepackage{float}  
\usepackage{enumitem}
\usepackage{tcolorbox} 
\usepackage[labelfont=bf]{caption}
\usepackage{subcaption}     % for subfigures
\usepackage{graphicx}

\title{Data Hunches: Incorporating Personal Knowledge into Visualizations}

\def \shortauthor {Lin et al.}
\author{
  Haihan Lin \\
  University of Utah
     \And
   Derya Akbaba \\
   University of Utah 
      \And
   Miriah Meyer \\
   Linköping University
      \And
      Alexander Lex \\
  University of Utah \\
  \texttt{alex@sci.utah.edu}
}

\begin{document}

\twocolumn[
  \begin{@twocolumnfalse}
    \maketitle
    
\begin{abstract}
The trouble with data is that it frequently provides only an imperfect representation of a phenomenon of interest. Experts who are familiar with their datasets will often make implicit, mental corrections when analyzing a dataset, or will be cautious not to be over-confident in any findings if caveats are present. However, the implicit knowledge about the caveats of a dataset are typically not collected in a structured way, which is problematic especially when teams work together who might have knowledge about different aspects of a dataset. In this work, we define such analyst's knowledge about datasets as \textit{data hunches}. We discuss the implications of data hunches and propose a set of techniques for recording and communicating data hunches through data visualization. Furthermore, we provide guidelines for designing visualizations that support recording and visualizing data hunches. We envision that data hunches will empower analysts to externalize their knowledge, facilitate collaboration and communication, and support the ability to learn from others' data hunches.
\end{abstract}
 
\vspace{2mm}    
    % keywords can be removed
\keywords{Data Visualization\and Uncertainty\and Situated Knowledge\and Data Hunches}
\vspace{8mm}
  \end{@twocolumnfalse}
]

\begin{tcolorbox}[floatplacement=!b,float,left=2mm,right=2mm,top=1mm,bottom=1mm]
\small
This is the authors' preprint version of this paper. License: CC-By Attribution 4.0 International. Please cite the following reference: \\
Haihan Lin, Derya Akbaba, Miriah Meyer, Alexander Lex. 
Data Hunches: Incorporating Personal Knowledge into Visualizations.
2021.
\end{tcolorbox}

\begin{figure*}
  \includegraphics[width=\textwidth]{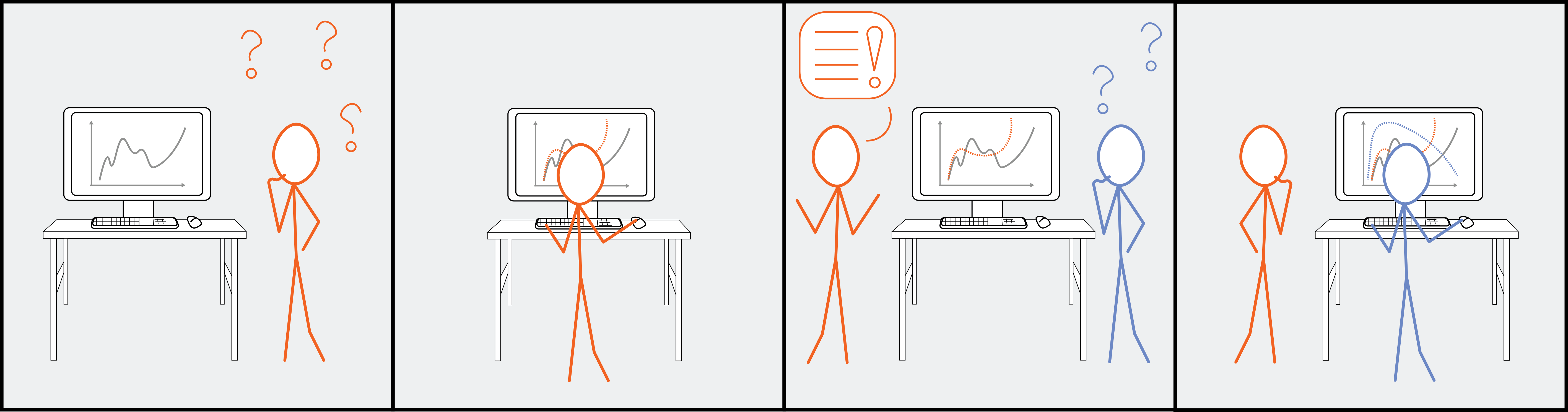}
  \caption{A representation of the process of recording and and collaborating with data hunches. When an analyst looks at a data visualization and has a hunch about the data, they are able to externalize the data hunch through visual methods that we introduce in this paper. Following the recording of their data hunch, others can review the hunches and potentially record their own.}
  \label{fig:teaser}
\end{figure*}

\section{Introduction}

%Data and data visualizations are ubiquitous today in scientific research as well as daily communications. We live in a data saturated era, and people make critical decisions based on data and analytical models of data. Data visualization has become a main tool for analyzing, interpreting and communicating data. As a new resource, 
While data-driven decision-making and reasoning is now considered the gold standard across many fields such as business~\cite{brynjolfsson_rapid_2016}, sports~\cite{troilo_perception_2016}, and science~\cite{nationalsciencefoundation_nsf's}, data workers are also aware of the incompleteness and imperfections of data~\cite{boukhelifa_how_2017, muller_how_2019}. In scientific data analysis, for example, caveats about the data are often captured in (digital or paper) lab notebooks, as comments in analysis scripts, as datasheets recorded alongside a dataset~\cite{gebru_datasheets_2018}, discussed with colleagues in lab meetings, and reported to a general audience in methods sections of scientific papers. 

These processes, however, detach a person's knowledge about caveats to the data from the tools in which the data is analyzed. For example, if a caveat is recorded in a graduate student's lab notebook, their successor who is asked to re-analyze the data may not have access to the notes and may misinterpret the data. Or, if a caveat is noted in the methods section of a paper, a casual reader might skip that section and potentially misinterpret a visualization of the data. 

We encountered this challenge in a design study where we collaborated with clinicians who analyze data about blood transfusions to improve patient outcomes and reduce the use of valuable blood-bank resources~\cite{lin_sanguine:_2021}. When we showed a clinical collaborator a prototype of our visualization tool, he expressed concern about what he saw in the data: that the amount of recorded recycled blood---a patient's own blood that is re-used during surgery---was much lower than he expected. His expectation came from his experience that almost all surgeons making extensive use of blood recycling; he had a hunch that the low blood recycling values were due to the data frequently not being recorded during surgeries. A second collaborator focusing on patient blood management was promoting a more data-driven approach in his workplace and worried that when clinicians saw the same discrepancies they would lose trust in the visualization tool. And we had no way of recording and communicating the clinicians' hunch in the visualization.

Just like in this example, we frequently find that interactions with a visualization tool triggers experts to express knowledge about problems with data, but that these tools leave few options for them to communicate that knowledge. This knowledge is personal and not available to others, a phenomenon reported in other design studies~\cite{mccurdy_framework_2019,nowak_designing_2020,panagiotidou_implicit_2021}. 
But what if experts could record their hunches directly in a visualization tool in a way that allowed others to interpret the data alongside their hunches? What if a visualization tool supported---and encouraged---the explicit incorporation of expert knowledge during data analysis?

In this paper we address these questions by proposing a new concept, \textit{data hunches}, to describe the knowledge that analysts bring to their data analysis process that augments and complements their perception of the data, and in turn becomes part of their interpretations as demonstrated in the cartoon-style teaser image. Data hunches are personal knowledge about caveats to the data, such as knowledge about implicit errors~\cite{mccurdy_framework_2019} or ambiguities~\cite{nowak_designing_2020}. We argue that conceptualizing data hunches provides a new perspective on how to design visual analysis tools for expert, collaborative settings that allows for the sharing of knowledge about the ways that data is imperfect, partial, and uncertain. This new perspective opens design possibilities for how we might build visualization tools that support both the recording and communicating of hunches.

To this end, our work includes three core contributions:
\begin{itemize}[nosep]
    \item \textbf{A conceptualization of data hunches:} We define the term \textit{data hunches} and discuss its relationship to uncertainty, elevating the role of personal knowledge in visual data analysis.
    \item \textbf{An exploration of ways to record and communicate data hunches:} We demonstrate how data hunches can be recorded and communicated within a visualization context using visual techniques to facilitate collaboration.
    \item \textbf{A set of design recommendations for data hunches:} We reflect on design and ethical considerations for supporting data hunches within visualization systems.
\end{itemize}
We speculate how recording and communicating data hunches could be implemented through a demo tool, available at \url{https://vdl.sci.utah.edu/data-hunch/}. We also offer a discussion about the potential of data hunches, for both good and bad outcomes. Despite open questions about how to design for data hunches in ethical and usable ways, we consider this work a step towards a wealth of productive opportunities for valuing and including personal knowledge in visual data analysis.

\section{Uncertainty}
Measurement errors, modeling assumptions, heterogeneous data recording methods, and missing context are just some of the ways in which values stored in a dataset are neither a perfect nor complete view of a phenomenon in the world. 
The visualization community has a long history of researching methods to characterize, quantify, and visualize the limitations of data under the heading of \textit{uncertainty}. Within this literature, explicit definitions for uncertainty are sparse, with the definitions that do exist covering a range of interpretations. 
For example, Hullman refers to uncertainty as ``the possibility that the observed data or model predictions could take on a set of possible values''~\cite{hullman_why_2020}, while Bonneau et al.'s review states that ``uncertainty is the lack of information''~\cite{bonneau_overview_2014}. Fields outside of visualization add additional interpretations of the term, such as Walker et al.'s general notion of uncertainty as ``any deviation from the unachievable ideal of completely deterministic knowledge of the relevant system''~\cite{walker_defining_2003}, to Covitt et al.'s experiential definition of uncertainty as the ``ways in which scientists recognize and analyze limits in their studies and conclusions''~\cite{covitt_untangling_2022}.

Researchers have openly acknowledged the ill-defined nature of uncertainty, such as Boukhelifa et al.'s statement that ``there is no unified single definition of uncertainty across all domains. The general consensus is that there are different meanings and that the term itself encapsulates many concepts''~\cite{boukhelifa_how_2017}. Brodlie et al.\ go further and argue that the lack of a clear, consensus definition has held the field back: ``the self-referential problem of uncertainty about uncertainty terminology has been a notable stumbling block in this avenue of inquiry''~\cite{brodlie_review_2012}. 

In this paper we do not attempt to rectify the uncertainty about uncertainty. Rather, we point out the difficulty of precisely positioning theoretical perspectives in relation to existing uncertainty literature. Despite this challenge, we situate data hunches within the visualization uncertainty literature because this body of work has focused on understanding and characterizing how people, visualizations, and imperfect and partial data come together. More specifically, the uncertainty literature focuses on the ways that data are limited representations of the world, and how people (can) become aware of these limitations. This is in contrast to other visualization research threads that focus on expert knowledge more broadly, such as work on insight that considers different types of expert knowledge that impacts insight generation~\cite{karer_insight_2021}.
In this section we briefly summarize the literature on quantitative and qualitative uncertainty, and argue that data hunches complement the existing work in this area.
 
\subsection{Quantitative Uncertainty}

Researchers provide a variety of characterizations of uncertainty through descriptions of the many ways that data can be uncertain. Potter et al.~\cite{potter_quantification_2012} use a characterization from computational sciences that describes \textit{epistemic} versus \textit{aleatoric} uncertainty, with the former describing the ways in which a lack of knowledge about and from the data induce a computationally unknowable uncertainty, and the latter encompassing data limitations that can be assumed and modeled statistically. 
Padilla et al.\ compare \textit{direct quantitative} uncertainties and \textit{indirect qualitative} uncertainties~\cite{padilla_uncertain_2021}.
In this framing, direct uncertainties are quantifiable expressions such as confidence intervals and probability distributions, while indirect uncertainties can only be expressed qualitatively.
Direct, quantifiable expressions of uncertainty are typically computed from sources of imperfections and partialities such as: limited data collection resources, such as not being able to sample every person in a population of interest; limited measurement capabilities, such as the precision of an instrument; or limited knowledge about the future, such as the unpredictability of forecasting weather~\cite{thomson_typology_2005}. 

The visualization community has historically focused on developing and testing methods for visualizing quantifiable uncertainty~\cite{padilla_powerful_2020, bonneau_overview_2014, potter_quantification_2012}.
Some approaches attempt to intuitively encode uncertainty through modifications of a data item's graphical mark using quantile dot plots~\cite{padilla_uncertain_2021}, glyphs~\cite{johnson_next_2003,wittenbrink_glyphs_1996,padilla_powerful_2020}, or value-suppressed color schemes~\cite{correll_value-suppressing_2018}. Other approaches have instead explored visual representations that directly display summary statistics~\cite{potter_visualizing_2010, correll_error_2014,potter_interactive_2012} or use animations showing hypothetical outcomes~\cite{hullman_hypothetical_2015}. 
Researchers have also developed uncertainty-specific evaluation techniques, such as eliciting users' internal models of probability distributions, recording the effects of uncertainty on decision-making, and assessing participant's sense of confidence after viewing uncertainty visualizations~\cite{hullman_pursuit_2019}.
% Standard methods for testing the effectiveness of quantitative uncertainty visualization techniques such as measuring the ease of interpreting presupposed information from the uncertainty visualizations and users' preferences with the visualization~\cite{hullman_pursuit_2019}.
% Additionally, the community has even focused on the best methods for evaluating the effectiveness of uncertainty visualization techniques, as described in a review article by Hullman et al.~\cite{hullman_pursuit_2019}.

The community's characterization of what types of uncertainty are quantifiable would seemingly place our blood-reuse example in Section 1 as something that could be quantified---a limitation of the measurement capabilities. In principle we could attempt to model and quantify this limitation, but any metric is likely to be grossly inaccurate because of the abstracted nature of the knowledge about the imperfections, a point raised by Thomson et al.: 
``In addition to uncertain measures, analysts are concerned with abstract uncertainties such as the credibility of a particular source or the completeness of a set of information. As the uncertainty becomes more abstract, it is more difficult to quantify, represent, and understand''~\cite{thomson_typology_2005}.
Instead, researchers adopt different perspectives for abstract sources of uncertainty---qualitative perspectives---that focus on the knowledge people have about the limitations of data.

\subsection{Qualitative Uncertainty}
Qualitative uncertainty---also referred to as indirect or epistemic uncertainty---has been described as ``the quality of knowledge concerning how effectively facts, numbers, or hypotheses represent reality''~\cite{padilla_uncertain_2021}.  Definitions of qualitative uncertainty make explicit reference to knowledge, shifting the emphasis from exploitable information about the data, to inaccessible subjective knowledge: ``Epistemic uncertainty generally, but not always, concerns past or present phenomena that we currently don’t know but could, at least in theory, know or establish''~\cite{vanderbles_communicating_2019}. In contrast to quantitative uncertainty, qualitative uncertainty is not easily quantified, and is generally conveyed through ``caveats about data''~\cite{vanderbles_communicating_2019}.

The sources of qualitative uncertainty stem from the same imperfections and partialities that metrics for quantifying uncertainty pertain to. Boukhelifa et al.~\cite{boukhelifa_how_2017} classify these sources as imperfect, messy, and missing \textit{data}; imperfect and limited \textit{models}; (approximate) digital representations of data in a visualization \textit{interface}; and \textit{cognitive} differences between interpretations from individual analysts. Most relevant for the work in this paper are sources of qualitative uncertainty from data, defined by McCurdy et al.\ as \textit{implicit error}: ``a type of measurement error that is inherent to a dataset but not explicitly recorded, yet is accounted for qualitatively by experts during analysis, based on their implicit domain knowledge''~\cite{mccurdy_framework_2019}. 

The predominate way that visualization designers encode qualitative expressions of uncertainty is through text-based annotations. In her interview study with visualization practitioners, Hullman reports that ``uncertainty as a qualitative expression of a gap in knowledge came up in most interviews with interviewees as well as several survey responses. 62\% of survey respondents had used text to warn their viewers of the potential for uncertainty in results''~\cite{hullman_why_2020}. Other approaches use visual approaches to communicate qualitative uncertainty, such as the use of perceptually imprecise visual encoding channels like sketchiness~\cite{boukhelifa_evaluating_2012} or glyphs~\cite{nowak_designing_2020}. 
A different approach taken in both the visualization and machine learning communities is to explicitly expose information about the data collection process, providing analysts with contextual information that allows them to incorporate personal knowledge about potential shortcomings of the data during their interpretation~\cite{panagiotidou_implicit_2021,gebru_datasheets_2018,arnold_factsheets:_2019}. 

Several visualization systems explore ways for recording expert knowledge about qualitative sources of uncertainty. For example, in their tool for supporting public health experts, McCurdy et al.~\cite{mccurdy_framework_2019} designed a template with structured questions that enabled the experts to record what they know about implicit errors in the data. The recorded results were marked on the visualizations with glyphs that provided annotations when clicked.
Similarly, Franke et al.~\cite{franke_confidence_2019} collected confidence about data sources from historians through a web interface. These varying levels of confidence were then presented alongside other data in a hierarchical tree view, representing the distribution of confidence along different dimensions of the data source in question.

These two examples demonstrate that experts \textit{know} about limitations of their data; that the sources of uncertainty can be known. This is in contrast to existing definitions of qualitative uncertainty that position it as something unknowable: ``epistemic uncertainty describes uncertainties due to lack of knowledge and limited data which could, in principle, be known, but in practice are not''~\cite{potter_quantification_2012}. This focus on the unknowable also appears in more general uncertainty characterizations~\cite{bonneau_overview_2014}.
This contradiction leads us to ask the question: unknown by whom? 

The work we present in this paper is a significant shift in how we think about designing for qualitative uncertainty. Data hunches frame the knowledge about sources of uncertainty away from the visualization tool designers, to the experts who conduct the analysis. The concept of data hunches is an explicit acknowledgement that many sources of qualitative uncertainty are in fact known---known to the experts who can articulate the knowledge when triggered by their interactions with a visualization tool. This shift complements the visualization community's perspectives on uncertainty by focusing on knowledge about sources of uncertainty, and thus opening up opportunities to design new ways that visualizations can support recording and communicating this knowledge during data analysis.

\section{Data Hunches}
\label{sec:data_hunches}
Our community's current framing of uncertainty implies that data workers and tool builders are responsible for identifying, characterizing, and quantifying sources of uncertainty from data. Data hunches instead acknowledge and incorporate experts who come to data analysis with deep knowledge about the limitations of their data. As a complementary perspective, data hunches focus on knowledge about data, capture that knowledge from a diverse set of stakeholders, and are embeddable in the analysis process (instead of the data curation pipeline). We argue that this perspective offers a breadth of new opportunities for recording and communicating data hunches in support of richer data analysis.

Through data hunches we elevate the role that personal knowledge of the data plays in the process of understanding and analyzing it. More precisely, \textbf{we define \textit{data hunches} as an analyst's knowledge about how and why the data is an imperfect and partial representation of the phenomena of interest.} These hunches can range from the abstract---expressing concern about the validity of the data set---to the concrete---expressing a numerical value that is closer to the phenomenon of interest than the measured data. The scope of a data hunch can be individual data points, a complete dataset, or anything in between. 
Data hunches emerge when an analyst interacts with the data, triggering reflection about the ways the data is imperfect and partial based upon their inherent knowledge about the data collection process, domain, and more.
A data hunch can be based on the missing context necessary to fully comprehend the phenomenon, discrepancies between a mental model and data, opinions on the quality of the data generation process, and so on. They are knowledge about sources of qualitative uncertainty.
During data analysis, data hunches influence an analyst's interpretation of the data, derived knowledge, and decisions made.

We argue that data hunches are prevalent, but often implicit, in data analysis, and as important as the data itself. The knowledge experts bring to data analysis is a vital component of data-driven decision-making~\cite{morgan_use_2014, karer_insight_2021}; however, such knowledge is often recorded outside of visual analysis tools. This disconnect then requires mental gymnastics on the part of an analyst to incorporate back into the data analysis process, if it is not overlooked completely. In this paper, we envision that a set of tools that fully utilize what visualization can offer will provide a more visual and intuitive representation of experts' knowledge in visualizations. Using data and data hunches in tandem supports a richer representation of a phenomenon, leading potentially to improved analysis. By acknowledging and naming data hunches, we aim to elevate the potential for personal knowledge to actively and explicitly contribute to data analysis.

We purposefully scope data hunches for use within collaborative, expert settings for both pragmatic and ethic reasons. Pragmatically, previous work on collaborative visualizations highlights the value of designing tools that support sharing of expert knowledge. In the context of collaborative analysis sessions, Walny et al.~\cite{walny_visual_2011} studied the use of data visualizations on whiteboards in corporate offices and found that visualizations as sketches promote team discussions. In another example of data science workflows that utilize computational notebooks Wang et al.~\cite{wang_how_2019} found that data scientists annotated screenshots of visualizations when collaborating as a way to communicate limitations of a tool. Ethically, scoping data hunches to collaborative expert settings reduces potential harm, which we discuss further in Sec~\ref{sec:discussion}.

By identifying data hunches as productive and insightful expert knowledge, we can re-interpret past work on collaborative visualization tools with this framing, like: a visualization designer can incorporate commenting and discussion features to promote externalization of data hunches~\cite{mccurdy_framework_2019,heer_voyagers_2009}; apply provenance tracking to record actions they took based on hunches to wrangle the data~\cite{feng_hindsight:_2017, cutler_trrack:_2020, gadhave_predicting_2021, gadhave_reusing_2022}; use visualization techniques like linked views and visualization states to show a collection of data hunches~\cite{viegas_manyeyes:_2007,mathisen_insideinsights:_2019,isenberg_collaborative_2011}; and much more. We see a wealth of opportunities for incorporating data hunches into old and new ways of visually analyzing data.

\begin{figure*}[t]
    \centering
    \includegraphics[width=\linewidth]{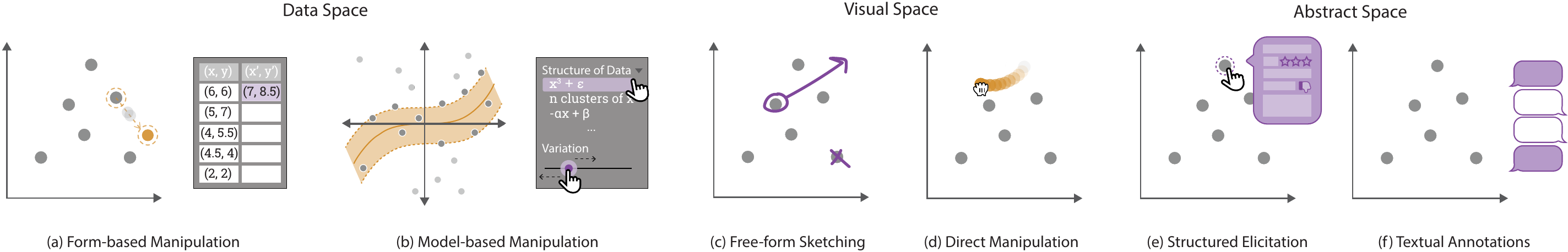}

    \caption{A set of abstract techniques to record data hunches. We distinguish across three spaces: data, visual, and abstract. Recording in \textbf{data space}, such as through (a) form-based manipulation and (b) model-based manipulation, affords a basic technique where analysts can externalize hunches either by manually inputting items, or expressing them through models. Recording in \textbf{visual space}, such as through (c) free-form sketching and (d) direct manipulation, affords visual recording of data hunches, ranging from sketching to dragging graphical elements to represent an analyst's hunch. Recording in \textbf{abstract space}, such as through (e) structured elicitation and (f) textual annotations, enables analysts to record hunches that cannot be recorded in the data and visual spaces.}

    \label{fig:externalize_design_space}
\end{figure*}

\section{Types of Data Hunches}
We identify three types of data hunches in support of determining suitable methods for reporting and communicating expert knowledge about limitations of data.

\paragraph{Assessment Hunches:} Assessment data hunches speak to the trustworthiness or quality of a dataset or individual data items, or simply provide context. Assessment hunches can take different forms, ranging from ratings (thumbs-up/down, numerical scores, etc.), to written comments about data items or datasets. 
    
\paragraph{Structural Hunches:} Structural data hunches state that certain data points or relationships should not be included in the dataset (exclusion) or that a data item or relationship is missing (inclusion). In a network dataset, for example, an inclusion data hunch could be used to indicate that an edge is missing. For many data types, inclusion hunches should also be combined with an estimate of a value of the included hunch. For example, when indicating that an item is missing from a dataset, the data hunch could also contain an estimate of the value of the item.  
    
\paragraph{Value Hunches:} Value hunches make a statement about how a specific data value should be different from what is recorded in the dataset. Value hunches equally apply to numerical, categorical, and textual/label data. For example, a value hunch for a category could state that an item should be in category A instead of category B. 
    
For practical reasons, we found it useful to further distinguish three methods of expressing value hunches, reflecting different levels of ``precision'' about a value hunch.  \textit{Directional Hunches} express that values should be different (higher or lower) from the recorded value without giving a specific value. They are a middle ground between assessment hunches, which make no statement about directionality, and hunches that give estimates for actual values. \textit{Specific Value Hunches} expresses exactly how values in a dataset should be different. For example, a value data hunch could state that the value encoded by a bar chart should be 20 instead of 12.
\textit{Value Range Hunches} acknowledge uncertainty about the value to be specified. Instead of expressing a specific value, it states that a value should be within a certain range. For example, a range hunch could state that the value of a bar should be between 18 and 22 instead of 12. We acknowledge that different sub-types of any of the higher-level types of data hunches could be useful in different contexts. 

All these types of data hunches can be expressed for an individual data item, groups of data items, or whole datasets. For example, a value data hunch could apply to a single point (this should be twice as much), to a few data points (all of these items should be twice as high), or to the whole dataset (all data points should be twice as high). These different types of data hunches also establish that hunches may not always be precise, supporting analysts to record some data hunches across varying levels of knowledge. In the next sections we discuss our methodology on how we design with data hunches in mind. With an emphasis on visual methods, knowledge about sources of qualitative uncertainty can be represented through graphical elements and interpreted context of the data.

\section{Methodology}
\label{sec:methodology}
Our methodology for theorizing about data hunches and developing a framework for recording and communicating data hunches was based on reflective practices~\cite{fleck_reflecting_2010, schon_reflective_2017}. We began by reflecting on our experiences working with a variety of domain experts who have rich knowledge about their data, knowledge that was not captured in their datasets. Through group discussions about our experiences, we recognized the missing formalization of personal knowledge and its impact in data analysis. We began mapping out the scope of data hunches, the relationship between data hunches and existing visualization concepts, and how hunches have been reported in the existing literature. This process included a literature search into data feminism, critical data studies, and uncertainty, as well as searching works on design studies and reviewing any reported sources of qualitative uncertainty in previous design studies.

After investigating the landscape of data hunches, we iteratively developed our understanding of data hunches. The iterations critically reflected illustrative examples from our prior experiences, and design spaces proposed for interactive visualization interfaces, uncertainty visualization, and collaborative sense-making. We additionally received feedback from our research lab and colleagues and made adjustments accordingly. We used the design space to re-imagine visualization systems presented in several design studies~\cite{heer_voyagers_2009,mccurdy_framework_2019,lin_sanguine:_2021}. 

Initially, we considered our framework for recording data hunches as a medium for collecting input and knowledge about a dataset from a general audience. Our reasoning was that the crowd might have insights about datasets based on their own experiences, such as collective and local knowledge about a COVID-19 dataset. We came to appreciate that a key challenge of data hunches, however, is that they could be used to explain away inconvenient data points, or that they could exacerbate the problem of confirmation bias~\cite{fischer_knowledge_2019,lee_viral_2021}. We thus made the decision to argue for scoping data hunches to collaborative settings with groups of experts who are supported by networks of trust~\cite{passi_trust_2018}. We discuss the potential benefits and harms that could be associated when data hunches are implemented for general audience systems in Section~\ref{sec:discussion}. 

Finally, we used an iterative design and development process for our prototype system, with the intent of embodying our ideas and evaluating their feasibility. We sketched alternative design ideas and implemented promising solutions, which we then evaluated within our team (see the supplementary file for examples). We designed and implemented many variations that we subsequently abandoned for various reasons. We describe the lessons learned from this process that stretched over more than a year as guidelines in Section~\ref{sec:guidelines}.

\section{Recording Data Hunches}
\label{sec:recording}

A key aspect of data hunches is that they are expressed during analysis by a diverse set of stakeholders. Hence, visualizations of the data are the ideal medium to express, record, and consume data hunches. 
In this section, we explore the set of approaches that can be used for recording data hunches on top of a visualization.

\begin{figure*}[t]
    \centering
    \includegraphics[width=\linewidth]{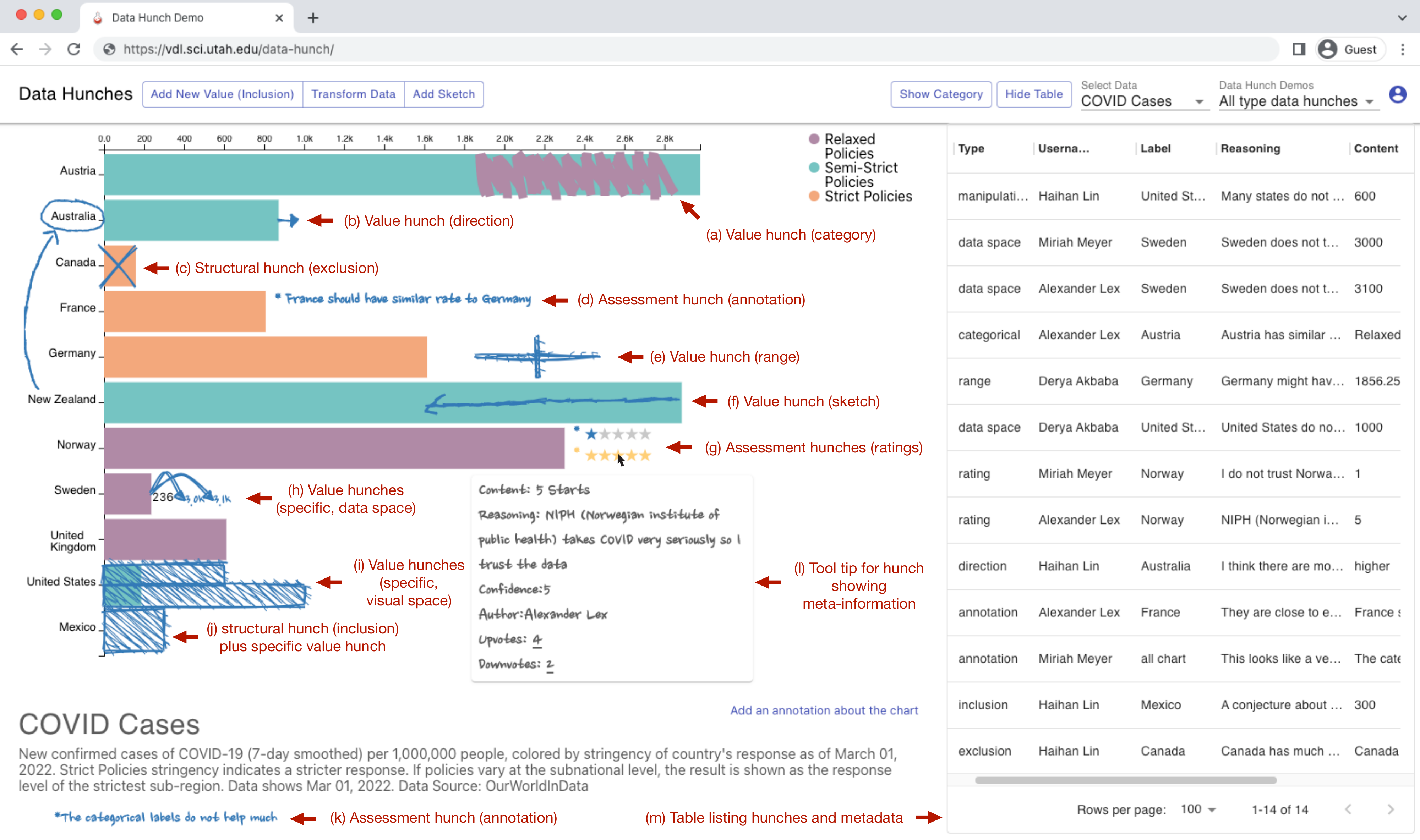}

    \caption{Our prototype showing a variety of data hunches expressed using tailored methods and designs. Note that red annotations have been added to explain the figure, everything else is as it appears in the prototype. The main chart, showing COVID-19 data and data hunches, is on the left. In the chart, we demonstrate (a) a categorical value hunch that indicates Austria should be colored in purple as it has relaxed policies which are not reflected in the data. (b) A directional value hunch expresses that the data value should be higher, without specifying details. (c) A structural hunch indicates that Canada should not be part of this chart. (d) An assessment hunch is used to comment on the values associated with France. (e) A range value hunch indicates that German values should be higher within a certain range. (f) A value hunch expressed using free-form sketching indicates that values for New Zealand should be lower. (g) Two assessment hunches show diverging perspectives about the quality of Norway's data. (h) Two value hunches expressed in data space indicate that values in Sweden should be much higher. They are rendered as arrows/labels as they would break the scale. (i) Two specific value hunches show alternative opinions on how high the values for the United States should be. (j) A structural hunch indicates that Mexico should be included in the chart and gives an estimate for a value. (k) An assessment hunch comments on the whole of the dataset. (l) A tool-tip shows information for a selected hunch, including reasoning, confidence, endorsements, and rejections. (m) A (hide-able) table lists all hunches. \href{https://vdl.sci.utah.edu/data-hunch/?data=COVIDData\&vol=2}{[View demo.]}}
    \label{fig:all_dh_prototype}

\end{figure*}

\subsection{Data Space}

%As analysts explore and conduct analysis on datasets, they often discover missing data, duplicates, or unusual patterns in the data. These hunches about the data comes to fruition without any visual aids, hence can be directly expressed through numbers and models. 
A basic method to record a data hunch is to manipulate the data in data space: before the data has been mapped to a visual element. We consider form-based manipulation and model-based manipulation as the two main methods for recording data hunches in data space. \textbf{Form-based manipulation}, shown in Figure~\ref{fig:externalize_design_space}a, is concerned with inputting a data value or an attribute of the data point through a form, a table, etc., and is suitable for data hunches for specific data items. \textbf{Model-based manipulation}, illustrated in Figure~\ref{fig:externalize_design_space}b, uses a model to bulk-input or edit data values, e.g., through a mathematical function. Note that manipulating the data in data space does not imply that the original data is overwritten. 

Previous works have explored ways to express models and values to record knowledge in visualizations. Marasoiu et al.~\cite{marasoiu_clarifying_2016}, for example, presented an interface that allows users to sketch models, which then generates data points based on the sketch, as a way to facilitate communication between customers and analysts. Romat et al.~\cite{romat_activeink:_2019} included data editing in their digital ink externalization system, a functionality requested by participants. Although this functionality was added post facto, it illustrates a preference for editing data directly in systems.

Data space is not well suited to communicate the data hunches that have been recorded in the context of visualizations, as e.g., a tabular representation of data hunches would be detached from the visualization of the data. Instead, designers will have to consider methods to visualize data hunches provided in data space in visual space.

\subsection{Visual Space}

Recording data hunches in visual space on top of a visualization provides a direct connection between a data hunch and the visualization. Analysts can think about the data hunch in visual terms as they manipulate it, and consider other data points that are visualized while recording their hunch. Another key benefit of recording data hunches in visual space is that, to a large extend, the same encodings can be used for recording and communicating data hunches. We suggest two techniques for recording data hunches in visual space: free-form sketching and direct manipulation.

\textbf{Free-form sketching}, shown in Figure~\ref{fig:externalize_design_space}c, refers to adding visual elements directly to a visualization, using approaches such as pen/mouse-based sketching, or adding elements to a visualization using functionality similar to a drawing program. The technique provides freedom for analysts to express their data hunch in the way they see fit and can record a variety of data hunches, including structural exclusion through crossing out data points, value/directionality (e.g., a value should be higher in reality) through drawing an arrow, categorical value through shading an area in a color, and value range through drawing an area where a value is expected to be. Visual markups have been used to for note taking and communicating though process in visualizations~\cite{kim_inking_2019, marasoiu_clarifying_2016,romat_activeink:_2019}. The process of graphical externalization also helps with the understanding of and reasoning about visual information~\cite{hegarty_individual_1997}. A downside of markup is that it cannot (easily) be converted into structured data, and hence is only connected to the visualization, but not to the underlying data, making re-use of these hunches in other visualizations of the same dataset exceedingly difficult. 

\textbf{Direct manipulation}, illustrated in Figure~\ref{fig:externalize_design_space}d, involves moving, resizing, removing, adding, or otherwise changing parts of the visualization that encode data. While restricting analysts to the marks and channels of the visualization, direct manipulations offer beneficial affordances: dragging a bar element is easier with a mouse than sketching a new bar, for example. Manipulated marks are also straight-forward to translate into data space. Previous works have suggested direct manipulation on visual encodings is a viable way to edit data and provide visual demonstrations of thought processes. Baudel~\cite{baudel_information_2006} presented editing single or groups of data items in a dataset using graphical manipulations in data visualizations. Saket et al.~\cite{saket_visualization_2017} used graphical manipulations (through repositioning, resizing, and recoloring marks in visualizations) to help users express their expected visualization with increments in direct manipulations, and in turn, the system suggests visual transformations. A drawback of direct manipulation is that each possible manipulation has to be designed and implemented for each chart type, in contrast to free-form sketching, which can be implemented once and re-used for all types of hunches and charts. 

\subsection{Abstract Space}
Assessment data hunches can only be expressed through text, comments, or ratings. For example, an analyst might know that a data source is unreliable, but might not have a concrete idea on what the true data should be. To record such a hunch, they want to add comments to the data and the data visualization. Such hunches are recorded in ``abstract space'', as they do not directly suggest a different structure or value. We identify two methods through which hunches can be recorded in abstract space: \textbf{structured elicitation} and \textbf{textual annotations} (in addition to e.g., hand-writing using free-form sketching features).

Structured elicitation (Figure~\ref{fig:externalize_design_space}e)  uses form-based UI elements, ratings, and up/down votes, while textual annotations (Figure~\ref{fig:externalize_design_space}f)  are free-formed notes. Previous works have explored the use of rating, structured form, and textual annotations in data visualizations. McCurdy et al.~\cite{mccurdy_framework_2019} used structured forms to elicit data hunches from domain experts, and Goyal et al.~\cite{goyal_effects_2013} offered more freedom to users by allowing them to use a notepad for free-form notes during their experiment. Structured elicitation is different from form-based manipulation in data space: while both can use forms, structured elicitation is about assessment, while form-based manipulation is used to express concrete hunches in data space.

\section{Guidelines for Designing for Data Hunches}
\label{sec:guidelines}

To demonstrate the feasibility of the techniques we proposed in Section~\ref{sec:recording} and to explore possibilities of communicating and recording data hunches, we developed a prototype, shown in Figure~\ref{fig:all_dh_prototype}, allowing users to record their data hunches on a simple bar chart. As described in Section~\ref{sec:methodology}, we used an iterative design process. We describe insights gleaned from this process for designing visualizations that support recording and displaying data hunches in the following in the form of design guidelines.

\paragraph{\textbf{Do not change the original data.}} 
Techniques to express and communicate data hunches aim to enable analysts to express their knowledge about the data, but not to alter the dataset. Data hunches and original data are different entities and should be clearly separated, to both avoid confusion about the difference between data and data hunch and to retain the integrity of the original data. Data hunches are also only valuable in the context of the recorded data that they refer to. Furthermore, while transparently expressing data hunches, such as doubts or ideas about what a data point should be, are valuable contributions to the analysis process, editing data can be considered deceptive or even fraudulent. 

From a technical aspect, designers should treat recorded data hunches as another dataset completely that only references (elements in) the original dataset. In our prototype, for example, data hunches are recorded as a separate dataset, which is shown in a table next to the visualization (Figure~\ref{fig:all_dh_prototype}m). 
An unfortunate consequence of separating hunches form data is that off-the-shelf visualization systems and libraries are unlikely to make it easy to render data hunches in addition to the underlying data.

\paragraph{\textbf{Make data hunches distinct.}} 
Consistent with our arguments about not changing the original data, data hunches also need to be clearly distinguishable from the original data within the visualizations. Furthermore, the encoding used for data hunches should not only be distinct from the primary encoding, but should clearly communicate that what is shown is not original data.

\begin{figure}[t]
    \centering
    \includegraphics[width=\linewidth]{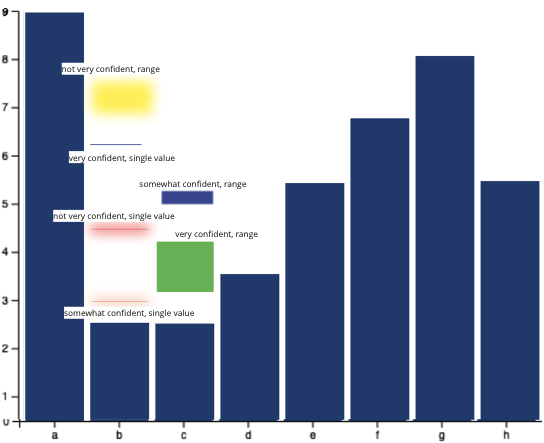}
 
    \caption{Original, non-sketchy design for data hunches. While the data hunches were distinguishable from the original data by color, the distinction was not immediately obvious and could be confused with additional data values that are part of the original data. Hence, we abandoned this design in favor of sketchy renderings.}

    \label{fig:non-sketchy-design}
\end{figure}

In our prototype, we use sketchy rendering~\cite{wood_sketchy_2012} for visual elements and handwriting-style fonts to make data hunches discernible from the crisp, clean lines of the original visualizations (see Figure~\ref{fig:all_dh_prototype}). The goal of using sketchiness was to make it obvious, even to first time users, that data hunches are not original data but that instead that they represent people's thoughts and knowledge. Wood et al., for example speculate that ``sketchy [..] visualization has a role to play in constructing visualization narratives where an author’s voice is important''~\cite{wood_sketchy_2012}.
In our first designs, shown in Figure~\ref{fig:non-sketchy-design}, we attempted to render data hunches using a different color, and experimented with gradients to communicate uncertainty and ranges. However, we abandoned this design as we realized that it could lead to confusion, as the marks could be interpreted as belonging to the original data. To avoid this confusion, we developed a rule for we applied throughout our design process that all data hunches should look as if they were hand-sketched or written on top of a visualization, to emphasize the humanness of the hunches. 

We do not argue that sketchiness is the only, or even the best choice to communicate data hunches. Other designs, tailored to other visualization techniques and affordances are conceivable. This reasoning applies to all of the implemented features we describe in this section.

Another consideration is to be mindful of how disruptive data hunch encodings could be when placed over the original visualizations. It is essential that the original visualizations remain visible to properly interpret data hunches. For example, in an early design, we rendered a bar representing a larger data hunch over the original bar. We abandoned this idea in favor of hatched bars which ensure that the underlying data point remains visible. 

% \begin{figure*}[t]
% \begin{subfigure}{0.32\linewidth}
%      \centering
%     \includegraphics[width=\linewidth]{figures/context_menu.png}
%     \subcaption{Context menu}
%     \label{fig:context_menu}
% \end{subfigure}
% \hfill
% \begin{subfigure}{0.32\linewidth}
%   \centering
%   \includegraphics[width=\linewidth]{figures/input.png}
%     \subcaption{Direct manipulation}
%     \label{fig:input}
% \end{subfigure}
% \hfill
% \begin{subfigure}{0.32\linewidth}
%      \centering
%     \includegraphics[width=\linewidth]{figures/in_chart_input.png}
%     \subcaption{Form-based input}
%     \label{fig:in_chart_input}
% \end{subfigure}

\paragraph{\textbf{But, make data hunches similar.}} 
Data hunches should use the same or similar visual encodings as the visualizations of the original data. While this seems like a direct contradiction of the guideline on making data hunches distinct, we believe it is important that data hunches and the original data can be read together without the need for mental conversion. For example, a value data hunch could be expressed as a written numerical value on top of a bar chart. However, we argue that this would make it difficult for an analyst to judge the relative differences between the original value and the data hunch. If both are expressed using the same visual channel (size/position in the case of a bar chart), comparisons are easier to do. Our prototype uses sketchy bars on top of regular bars for numerical specific value hunches (Figure~\ref{fig:all_dh_prototype}i), and hatched color for categorical hunches (Figure~\ref{fig:all_dh_prototype}a), in both cases using the same encoding channel as the original data. However, some hunches cannot be expressed using the same visual encoding. For example, a range value hunch (providing an estimate that a value should be in a certain range) is not compatible with a bar chart encoding used for the original data. To address this, we use a position encoding on the same scale as the bars, showing the middle and the extend of the range (Figure~\ref{fig:all_dh_prototype}e). We use similar techniques for directional value hunches (Figure~\ref{fig:all_dh_prototype}b) and hunches that do not fit on the scale of a chart (Figure~\ref{fig:all_dh_prototype}h). 

To reconcile this guideline with the guideline of making hunches distinct from visualizations of the original data, in our demo we relied on using an additional visual channel that was not used in the original visualization: sketchy texture.

\paragraph{\textbf{Keep data hunches close.}}
As assessment data hunches (comments, ratings, etc.) are not expressed in data space, using the same visual encoding is not feasible. However, designers should ensure that assessment hunches can still be read easily together with the original data. For example, instead of showing assessment hunches in a table, they could be rendered next to the element they are referring to, illustrated for annotations in Figure~\ref{fig:all_dh_prototype}d and for ratings in Figure~\ref{fig:all_dh_prototype}g. If a textual hunch requires more space than is available in the chart, we truncate the comment and reveal the full text in a tooltip. We also considered what to do with assessment hunches referring to the whole dataset and opted for placing a note and an asterisk next to the chart title (Figure~\ref{fig:all_dh_prototype}k); our reasoning is that analysts might read the title and caption as they are attempting to understand the visualization.

\begin{figure}[t]
\begin{subfigure}{0.48\linewidth}
  \centering
  \includegraphics[width=\linewidth]{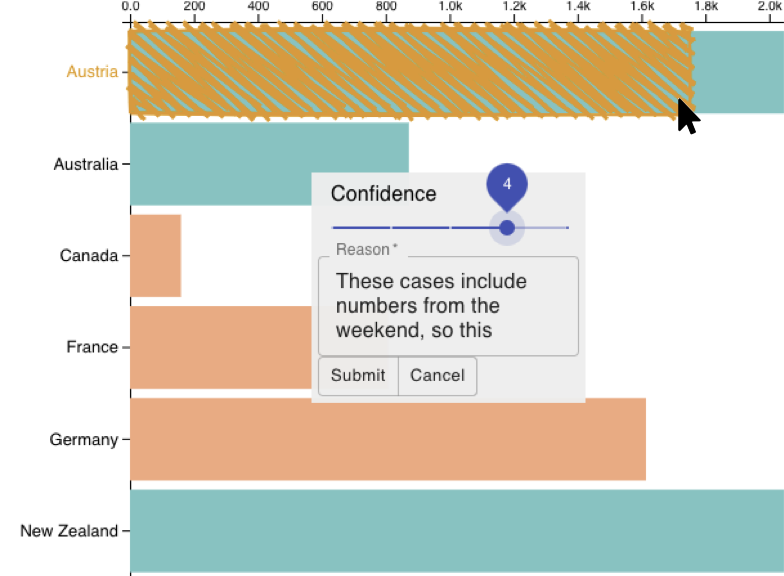}
    \subcaption{Direct manipulation}
    \label{fig:input}
\end{subfigure}
\hfill
\begin{subfigure}{0.48\linewidth}
     \centering
    \includegraphics[width=\linewidth]{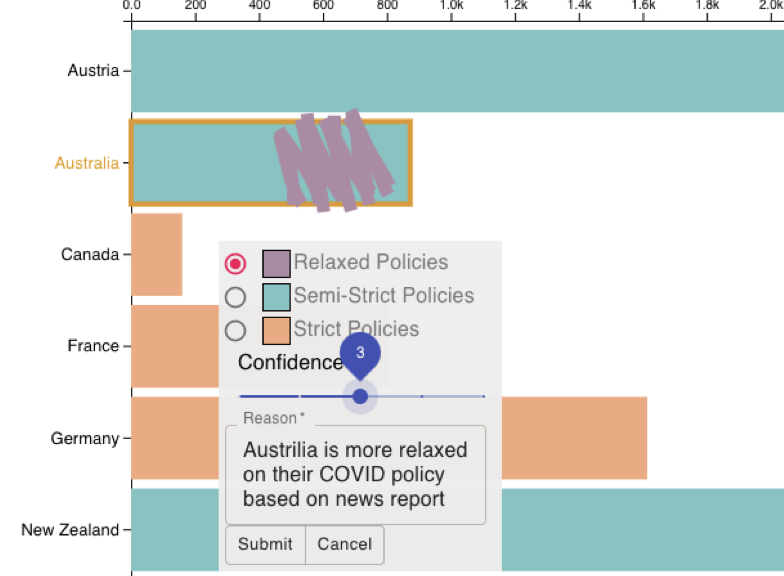}
    \subcaption{Form-based input}
    \label{fig:in_chart_input}
\end{subfigure}
 
\caption{Recording methods for data hunches, reasoning and confidence. (a) Recording a data hunch using direct manipulation of a bar. A confidence and reason is also elicited through a form. (b) Form-based input for a categorical hunch. Selecting an option immediately shows a preview. The input forms are placed close to the selected data point for clear association with the data point.}
\label{fig:input_methods}

\end{figure}

\paragraph{\textbf{Use direct manipulation.}}
Data hunches emerge when analysts explore the data and examine the data visualizations, hence, the thinking and analysis process happens in visual and data space. While we lay out different approaches for recording data hunches in Section~\ref{sec:recording}, we argue that, recording of data hunches should be done as close to the way the data and the data hunch is presented as possible. In our prototype, we trigger recordings by right-clicking on a mark or legend whenever possible and provide methods to record a data hunch through direct manipulation of the data hunch as it will appear once it is recorded (Figure~\ref{fig:input}).  
When recording a hunch in data space, or when recording an assessment hunch, a visualization can provide visual feedback for the analysts (Figure \ref{fig:in_chart_input}). Also, in our prototype, we place the input forms for data hunches right next to the visual elements in the chart.

\paragraph{\textbf{Design for data hunches.}}
Not all visualization techniques are equally well suited to visualize data hunches. In our prototype, we have chosen a bar chart with categorical values because bar charts are an important class of visualizations, and because they have affordances that are well compatible with data hunches. For example, an analyst could express a value data hunch on top of an individual bar without affecting a neighboring bar. Other visualization techniques, however, do not equally support such similarities. For example resizing a segment in a pie chart, or in a stacked bar chart, requires affecting the other data marks, or overplotting. 
%While different approaches for designing data hunches for bar charts may exist (small multiples, for example), they are likely not as effective at integrating data hunches with the original data in one visualization. 

Another consideration is the complexity of a chart: the more complex, and the more visual channels are used, the more difficult it will likely be to find a suitable design for data hunches. Similarly, visualizations that give overviews of large amounts of data, like cluster heatmaps, will require different approaches for data hunches, as the manipulation of individual data items is less relevant and the visualizing of the hunches more challenging.  

But even when using a visualization technique that is well suited for data hunches, like bar charts, line charts, scatterplots (see Figure~\ref{fig:externalize_design_space}), node-link diagrams, or tabular visualizations, there are certain things we recommend that a visualization designer keep in mind to better support data hunches. 
For example, our original design used vertical bars (see Figure~\ref{fig:non-sketchy-design}). However, we quickly found that vertical bars are problematic for rendering longer comments (assessment hunches) next to the bar due to the text orientation, so we switched to horizontal bars. Likewise, our original design had a chart title and a subtitle at the top, which made it difficult to find a suitable place for comments on the whole chart. Hence, we moved the title below the chart, and reserved space below the subtitle for comments. Finally, we found it useful to leave white-space from the beginning, so that data hunches can be easily expressed and rendered. For example, our prototype has large margins to the right of the bar chart, so that larger value hunches and comments can be effectively rendered. While these specific examples may not directly translate to other visualization techniques, the larger lesson of thinking about position, layout, and space for data hunches as a designer creates a visualization, holds.

\paragraph{\textbf{Design for collaboration.}}
Data hunches are predominantly a medium to communicate knowledge about data to others, hence data hunches are inherently collaborative. It is also very common that data analysis activities are collaborative efforts in the first place. We argue that data hunches should be designed with collaboration in mind.  

\begin{figure}[t]
    \centering
    \includegraphics[width=\linewidth]{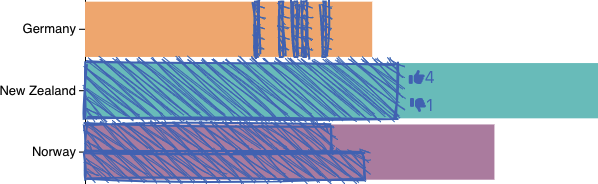}

    \caption{Collaboration and scalability. Collaboration features, such as endorsing, rejecting, and commenting on data hunches help in reducing the need for logging similar data hunches in the first place. Different visual encodings account for scalability in value data hunches. If more than three data hunches are logged for a single bar, we replace the sketchy bars with sketchy ticks.}
 
    \label{fig:scalability}
\end{figure}

Our prototype acknowledges this by allowing multiple people to log in and review data hunches and the underlying data. However, we also speculate that enabling multiple collaborators to only record data hunches might be insufficient, as collaborators might also want to endorse, reject, or comment on other's data hunches. To address this, we introduce features to endorse or reject a particular data hunch, using a thumbs-up or thumbs-down metaphor, illustrated in Figure~\ref{fig:scalability}. Another, more expressive method is to add capability to comment on data hunches. This way, a team can express their sentiment about a data hunch without having to re-specify a data hunch they endorse.

\paragraph{\textbf{Elicit context and accountability.}}
\textit{What} a data hunch says about the data is different from \textit{why} a person has the hunch. The context of a data hunch is as critical for its interpretation as the context of the data. 
% Scholars in the field of critical data studies have asserted that data cannot be considered out of context~\cite{klein_data_2020,seaver_nice_2015,boyd_critical_2012}, and identity can affect how trust is established in collaborations~\cite{zhang_identity_2009}. 
Similarly, the reasoning and identity of the author can effect how a data hunch is perceived and trusted. 

We propose that data hunches require reasoning for their meaning to be clear for context, and identity to establish trustworthiness of the hunch. As designers work with stakeholders to determine how data hunches are recorded, they should also explore how context can be recorded and shared.
%, as understanding the context of a data hunch and knowing the identity of the author is essential to evaluate trustworthiness. 
In our prototype, we require analysts provide reasoning for and express their confidence in a data hunch when recording it (see Figure~\ref{fig:input}). Additionally, the identity of the author is recorded as an attribute of the data hunch. We then visualize these attributes in a tool-tip (Figure~\ref{fig:all_dh_prototype}), although other, more salient approaches are conceivable. For example, it might be worth exploring the use of opacity to encode the confidence of a data hunch. These attributes not only enrich the recording of data hunches, they also allow for features such as filtering and sorting. 

We acknowledge a tension between revealing identities, to ensure accountability and leverage networks of trust, and the desire to be anonymous to record inconvenient opinions or facts. In prior work, for example, we found that experts in an organization were unwilling to record hunches under their name due to tensions in the organization. 
In addition to logged-in recording of data hunches, our prototype also provides a ``log-in as guest'' option to record a hunch without revealing one's identity. We further discuss this issue in Section~\ref{sec:discussion}.

\begin{figure*}[t]
    \centering
    \includegraphics[width=\linewidth]{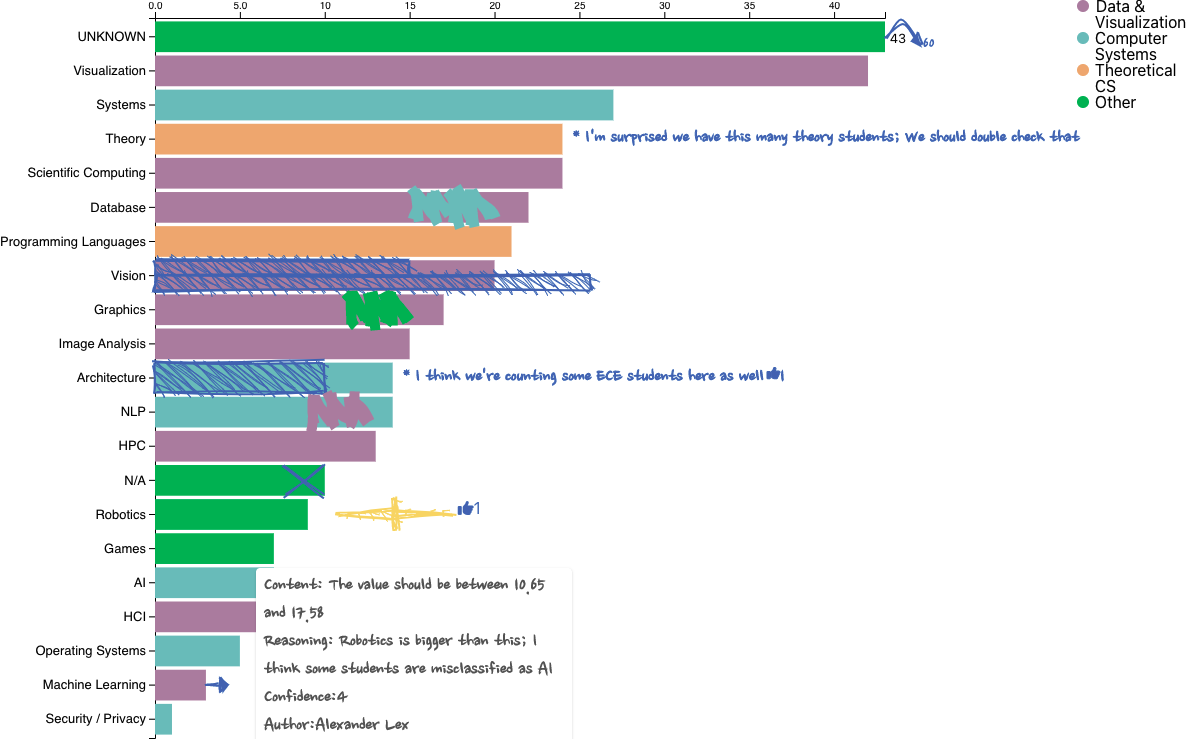}

    \caption{A usage scenario showing the data hunches of two faculty members about the size of research areas in a CS department.  \href{https://vdl.sci.utah.edu/data-hunch/?data=student\&vol=1}{[View Demo]}}
    \label{fig:case_study_students}

\end{figure*}

\section{Prototype Implementation}
\label{sec:prototype}

Our web-based prototype (available at \url{https://vdl.sci.utah.edu/data-hunch/}) implements  visualization methods for all types of hunches we describe (see Section~\ref{sec:data_hunches}). For specific value hunches, we also provide dedicated methods for larger numbers of data hunches. The prototype also implements all appropriate methods for recording data hunches (see Section~\ref{sec:recording}) for each type. A specific hunch, for example, can be recorded in data space through form-based manipulation, through model-based manipulation, as well as in visual space through direct manipulation and free-form sketching. 

A concern for designing hunches was scalability: how can we show many different data hunches reliably on top of a visualization. We address scalability in two ways: first, we implemented dedicated encodings for hunches in case to many hunches are recorded. Figure~\ref{fig:scalability} shows specific value hunches on top of three bars. We use sketchy bars for one or two hunches on top of a bar, but switch to sketchy ticks for more hunches, as shown on the top. However, switching from bars to ticks violates the guideline for making hunches appear similar to the original marks and hence is a trade-off we have to make.

Another aspect designed to help with scalability are collaborative features. If a data hunch is already present, another person sharing the hunch could endorse, reject, or comment on data hunches instead of logging a new data hunch, thereby reducing the number of hunches that need to be visualized simultaneously.

We use three example data sets: COVID-19 case counts in selected countries, greenhouse gas emissions across the food supply chain (both downloaded from OurWorldInData), and the size of research areas at the School of Computing at the University of Utah. Data is loaded from CSV files, while data hunches are stored in a Firebase database. Data hunches can be freely added by guests or after signing-in via Google. A tabular visualization (Figure~\ref{fig:all_dh_prototype}) gives an overview of all data hunches and their meta-data, though all relevant information about data hunches is also available through the visualization interface. 

Our prototype is open source; the code is available at \url{https://github.com/visdesignlab/data-hunches-package}. We use React and D3 for rendering the UI and the visualizations, and the RoughJS library (\url{https://roughjs.com/}), which is based on the techniques for sketchy rendering developed by Wood et al.~\cite{wood_sketchy_2012}, to render visual elements for data hunches in sketchy patterns.

\section{Usage Scenario}
\label{sec:usage_scenario}

Here we illustrate how data hunches could be used in a scenario with real data about the size of research areas in the University of Utah's School of Computing. This usage scenario is shown in Fig~\ref{fig:case_study_students}.
To begin, a faculty member first noticed was that while the largest bar was for \textit{UNKNOWN} students, there was another bar labeled \textit{N/A}. He recorded a hunch that \textit{N/A} should be removed (a structural hunch) and that \textit{UNKNOWN} should be bigger (a value hunch). 

Upon reviewing the classification of research areas into larger fields, he expressed concern that \textit{Databases} is classified as \textit{Data and Visualization}, given his knowledge about the type of research conducted by the database groups at Utah, and that it should rather be in the \textit{Computer Systems} category --- he recorded his hunch about the different classification of the bar (a value hunch). He expressed a similar data hunch for \textit{Graphics}, which should be in the \textit{Other} bin.

When reviewing the number of \textit{Architecture} researchers, he was surprised by the seemingly high number. He provided a value hunch that he considered closer to reality, and added a comment speculating that some ECE students advised by CS faculty are included in this count. Finally, he realized that \textit{Robotics} is likely shown smaller than it is, probably because some students are incorrectly classified as \textit{AI} instead. He left a value range hunch also noting his reasoning.

This faculty member then passed on the visualization to a second faculty member who reviewed his hunches, upvoted several, and left some additional hunches about her own thoughts on where the data did not reflect the make-up of the department. 

\section{Discussion and Future Work}
\label{sec:discussion}

% As we reflected on the process of defining data hunches, proposing guidelines for supporting them in visualizations, and experimenting with different techniques in our prototype, we made assumptions and design decisions to ensure their proper usage. 
Inspired by the breadth of opportunities that data hunches open up, in this section we present a series of discussion threads and future work possibilities. Some of these threads reflect on ethical considerations for ensuring data hunches are used in productive and positive ways, and others consider a number of technical and design challenges for consideration. We end with a brief statement about the opportunity that data hunches point to for considering new perspectives on knowledge.
% In this section, we discuss the challenges and the future direction for data hunches, and provide a discussion on the epistemology of knowledge on data. 

\paragraph{Challenges for Designing with Data Hunches.}
As a proof of concept, we used bar chart to demonstrate how data hunches can be implemented in data visualizations. As we developed the prototype, we realized how quickly the additional layers of data hunches can complicate the visualizations. It was challenging to add data hunches in the chart while keeping the original visualization legible. %Just like uncertainty visualizations, supporting data hunches can be a challenging tasks for visualization designers. 
For data hunches to become established, we need to develop designs for a wide range of visualization types. We believe that our guidelines apply broadly, but the specific design decisions in our prototype might not easily translate to certain types of space-filling visualizations, such as pie charts, treemaps, or icicle plots. Also, while our framework allows for structural hunches through inclusion and exclusion, tabular data is a simple case compared to network data, which has much more complex structural (topological) information. Hence, we believe that a good amount of design work remains to be done to make data hunches work well with such datasets.

\paragraph{Potential for Harm.}
In our advocacy for data hunches, we only focus on the use cases for analysts with rich knowledge about their field. This ensures that the data hunches are based on analysts' knowledge and experience of the field, and can provide a richer view of the phenomenon that the data represents. The limitation can also avoid misuse and misinformation in data hunches. In an ideal world, our definition of data hunches implicitly assumes that all users are positively contributing to the visualizations when expressing their opinions and knowledge of the data. However, even with good intentions, there is a risk that data hunches could be used to explain away inconvenient data points, or to reinforce an analyst's preconceived ideas. This is why it is critical that data hunches are treated fundamentally different from data, even if they are recorded in the same space. A design for data hunches should go to great lengths to avoid any confusion between the data and the data hunch. This is also why it is very important that data hunches come with explanations and justifications. Using these techniques, analysts can evaluate a data hunch holistically and judge whether it is reasonable.% \todo{say something about it would be interesting to study whether data hunches in the field are actually used this way?} 

A lot remains to be explored when considering how to support the public with the ability of recording data hunches on visualizations. Data hunches can encourage open conversations about data and visualizations, similar to the current online Q\&A forums. Previous research concluded that identity-based trust, social feedback, and exposure all have positive effects on knowledge contribution~\cite{guan_knowledge_2018}. Visualizations supporting data hunches are similar to online forums, where users can communicate their knowledge or opinions about the data through a set of techniques and receive feedback (such as upvotes, downvotes, and comments) from other users in the community, and the association with identity, feedback, and exposure can positively promote conversation and knowledge sharing about the data. However, what if a hunch is wrong? Or worse, what if a hunch is maliciously intended as misinformation? Previous research has discovered that data can be used in contradicting ways, depending on how people use data~\cite{lee_viral_2021}. It is possible that a system that supports recording and visualization of data hunches can be used with malicious intent to fulfil a personal agenda. Moderators, allowing voting, and providing reports mechanism can potentially help with the issue, but it remains an open question that requires further investigations. 

\paragraph{Trust and Privacy}
While we suggest that identities behind data hunches can play a big role in building trust for data hunches, privacy can become an issue and concern for analysts. In some settings, the politics of an organization or field could cause vulnerable people to remain silent about contradictory hunches, depriving others of important perspectives~\cite{mohammed_understanding_2001}. But anonymity can be equally caustic by invoking negative behavior toward others with opposite views~\cite{barlett_anonymously_2015}. How to ensure the value, credibility, trustworthiness, and transparency of data hunches is an important, yet open, question.

Another interesting, open question is: What happens to someone's trust in a visualization and the underlying data when data hunches are communicated in a tool? Previous work~\cite{kim_data_2018} has reported that \textit{social information} can affect a user's trust and memorability about the data visualization. We anticipate similar effects with the inclusion of data hunches. We argue in this paper that data is an imperfect representation of reality, and making that imperfection visible is one goal of our work. However, if data hunches make people less trusting, will designers avoid including them, as they sometimes do with uncertainty~\cite{hullman_why_2020}?

A reader may trust the visualization more when data hunches are provided by experts, or when data hunches are highly rated. On the other hand, if too many data hunches disagree with the original data, the reader may trust the source of the visualization less. In the end, the goal of conceptualizing data hunches and proposing a design space for them is to formally recognize the role of personal knowledge in understanding data and empower users to express their views. Designers should fully consider the possible impacts of data hunches before committing to including or excluding them. The work we present in this paper is only the first step in exploring a rich space about how, why, and when to include personal knowledge about data in visualizations.  

\paragraph{Application Scenarios}
Another important consideration is what types of visualization systems and scenarios are most appropriate for implementing mechanisms that support data hunches. We believe that most visual data analysis involves hunches, but designing and developing tools that support externalization and communication of multiple data hunches is likely to require significant effort. We see two possible application scenarios that we think justify this effort. On the one hand, data hunches could be integrated in widely used off-the-shelf visualization libraries. For example, adding the capability to visualize and record data hunches to a library such as Altair~\cite{vanderplas_altair:_2018}, an interactive visualization library for Python and Jupyter notebooks, could make data hunches accessible to a wide range of audiences. A second possible application scenario are bespoke systems that support recording and communicating data hunches to be implemented primarily in scenarios where the topic of the data is of shared interest among larger communities of experts. For example, a recent project elicited feedback from the scientific community on an animation of the SARS-CoV-2 protein structure~\cite{iwasa_sars-cov-2_2020}. Unlike visualization tools designed for an individual research lab, such applications target a wider audience with shared interests, where visualizing data hunches can lead to deeper impact compared to casual visualizations. Finally, as designs for data hunches become more common and libraries to add data hunches to visualization become available, it might be feasible to integrate recording and visualization of data hunches into a wider set of visualization tools.

\paragraph{Data Hunches as Structured Data}
Depending on the type of hunch and the method for recording it, data hunches can also themselves become structured data. Value hunches recorded via data space or direct manipulation, for example, are either directly available in the same space as the original data, or can be easily translated back into data space. They are hence different from e.g., annotations provided on top of a figure, as they can be re-used whenever a dataset is re-used. For example, if a data hunch is recorded for a visualization of a dataset in a Jupyter notebook, the data hunch could be shown not only in that one visualization, but could be propagated to all subsequent and prior visualization that are based on that dataset, thereby surfacing the hunch at all stages of the analysis process. Data hunches could also be preserved as datasets are updated, as long as the structure is preserved. For example, if a dataset is refined over time, possibly because of discrepancies recorded as data hunches, a new version of a dataset could be overlaid with data hunches recorded for the old version of the dataset, to see whether the hunches expressed still apply. This could be combined with an explicit comparison of datasets~\cite{gadhave_reusing_2022}. However, such a workflow would incur additional visual complexity and hence would require dedicated methods to manage that complexity. 

%An analysts can also use the proposed structure to record information about the data that does not necessarily speak to the quality, the information is categorized as metadata. In comparison, a data hunch is personal knowledge about data. Rather than influencing the interpretation of data, metadata provides instructions on how to interpret the data~\cite{gartner_what_2016}. Existing works call for disclosure and transparency on metadata~\cite{arnold_factsheets:_2019,gebru_datasheets_2018}, and an infrastructure for data hunches can potentially facilitate the functionality called for in those efforts.

\paragraph{Epistemology}
The visualization community has characterized the imperfect and partiality of data as uncertainty; however, this is not the only available classification. Within the fields of critical data studies and digital humanities, data is an artifact of decisions and situated contexts that reflect one captured slice of reality: ``\textit{Data are capta}, taken not given, constructed as an interpretation of the phenomenal world, not inherent in it''~\cite{drucker_dhq:_2011}. Data then, as an object of decision-making practices, is one representation of many possibilities. This recognition is important because it highlights the non-neutrality of data~\cite{dignazio_feminist_2016, correll_ethical_2019, dork_critical_2013}.
% these perspectives come from a different epistemological perspective, we speculate these epistemologies would be really interesting and productive to theorize around data hunches; we're curious whether other epistemological foundations would be more productive to consider --- how could we better characterize and describe relationships between data, vis, perfect and partial data 
These perspective stem from different epistemological roots. Feminist perspectives on data are based on the theory of \textit{situated knowledges}~\cite{haraway_situated_1988}. This theory posits that knowledge cannot be obtained from a single source, but rather is best derived through a collection and collaboration across partial and overlapping perspectives. Data, similarly, cannot fully represent the natural world. From this theoretical grounding, data and data hunches would capture different perspectives of reality but contribute to a richer more complete picture. This is only a musing though, leading us to bigger questions like: In what ways could other epistemologies be productive in visualization research? Or, could we better characterize and describe the entangled relationship between data, visualizations, and people?

\section{Conclusion}

In this work, we framed the personal knowledge about how representative data is, defining such knowledge as data hunches, and analyzed the implication of supporting data hunches in data analysis. We proposed techniques for recording and communicating data hunches in data visualizations, listed a design guidelines, and implemented a prototype to demonstrate data hunches in action. The ultimate goal of this work is to formalize and recognize the significant role that personal knowledge has in understanding data, which many works overlook, and elevate this personal knowledge into another form of information that can be explicitly recorded and utilized. Through this work, we seek to question the notion of data being the gold standard of representing phenomena in the world, and open up the potential to grow visualization research beyond constrained notions of data.

\section{Acknowledgements}

We wish to thank Anders Ynnerman, Ben Shneiderman, Ryan Metcalf, and the Visualization Design Lab for fruitful discussions and feedback. This work was supported by the National Science Foundation (OAC 1835904, IIS 1751238), ARUP Laboratories, and by the Wallenberg AI, Autonomous Systems and Software Program (WASP) funded by the Knut and Alice Wallenberg Foundation.

\bibliographystyle{unsrt}  

\bibliography{2021_data_hunch}  %%% Remove comment to use the external .bib file (using bibtex).

%\end{multicols}
\end{document}